\documentclass[a4paper]{article}

\usepackage{amsfonts}

\def\be{\begin{equation}}
\def\ee{\end{equation}}
\def\bea{\begin{eqnarray}}
\def\eea{\end{eqnarray}}
\def\({\left(}
\def\){\right)}
\def\<{\left<}
\def\>{\right>}

\def\[{\left[}
\def\]{\right]}
\def\tr{{\mbox{tr}}}
\def\be{\begin{equation}}
\def\ee{\end{equation}}
\def\bea{\begin{eqnarray}}
\def\eea{\end{eqnarray}}
\def\({\left(}
\def\){\right)}
\def\<{\left<}
\def\>{\right>}

\def\[{\left[}
\def\]{\right]}

\def\+{\bar}

\def\tr{{\mbox{tr}}}
\def\Tr{{\mbox{Tr}}}
\def\D{{\cal{D}}}
\def\t{\tilde}

\begin{document}
\pagestyle{empty}
\vskip-10pt
\vskip-10pt
\hfill %{\tt hep-th/yymmnnn}
\begin{center}
\vskip 3truecm
{\Large \bf
In quest of a generalized Callias index theorem}\\ 
\vskip 2truecm
{\large \bf
Andreas Gustavsson}\footnote{a.r.gustavsson@swipnet.se}\\
\vskip 1truecm
{\it F\"{o}rstamajgatan 24,\\
S-415 10 
G\"{o}teborg, Sweden}\\
\end{center}
\vskip 2truecm
{\abstract{We give a prescription for how to compute the Callias index, using as regulator an exponential function. We find agreement with old results in all odd dimensions. We show that the problem of computing the dimension of the moduli space of self-dual strings can be formulated as an index problem in even-dimensional (loop-)space. We think that the regulator used in this Letter can be applied to this index problem.}}

\vfill 
\vskip4pt
\eject
\pagestyle{plain}
\section{Introduction}
We do not know what six-dimensional $(2,0)$ theory really is. It is believed that it can sustain solitonic self-dual strings \cite{Lambert}, although no one today knows what a (non-Abelian) self-dual string really is. But if we break the gauge group maximally to $U(1)^r$, then we should be able to define the charges of these mysterious self-dual strings by the asymptotic behaviour of the $U(1)$ gauge fields. One should expect these asymptotic $U(1)$ fields to be (at least isomorphic with) a copy of the familiar abelian two-form gauge potentials (with self-dual field strengths).

It now seems to make sense to ask a question like, what is the dimension of the moduli space of self-dual strings of a given charge?

If the gauge group is $SU(2)$ and is broken to $U(1)$ by the Higgs vacuum expectation value (that should also determine the tension of the string), then the intuitive answer to this question is $4N$ where $N$ is the $U(1)$ charge in a suitable normalization, such that $N=1$ corresponds to one self-dual string. One may argue that half the supersymmetry is broken by the string. Therefore one string should sustain $4$ fermionic zero modes. Since some (half) of the supersymmety is unbroken there should also be $4$ corresponding bosonic zero modes. These are naturally identified with the translational zero modes associated with the four transverse directions to the string. Furthermore, the strings being BPS, should be possible to separate at no cost of energy (thus staying in the moduli space approximation). If we take them far from each other, one may suspect that we can just add $4$ bosonic zero modes from each string, to get $4N$ bosonic zero modes in total in a configuration of $N$ strings \cite{Harvey}.

It would of course be nice to have a proof of this conjecture. Could it be proven if one had some index theorem? We will not provide a full solution to this problem in this Letter. But we will make it plausible that the problem can indeed be solved by computing the index of a certain Dirac operator in loop space.

To address our index problem, we think that one can lend the methods that Callias \cite{Callias} used to prove his index theorem in odd-dimensional spaces. In our case we have an even number of dimensions (namely the four transverse direction) so it is apparent that we would have to construct a new type of index. This we do in section \ref{selfdual}. 

In section \ref{callias} we recall the Callias method \cite{Callias} to address index problems in open spaces, though we will modify Callias' regularization, using the more convergent exponential function to obtain the index, as the limit
\bea
\lim_{s\rightarrow \infty} \Tr\(\gamma e^{-sD^2}\),
\eea
(here $D^2>0$ and $\gamma = $diag $(1,-1)$) rather than
\bea
\lim_{M\rightarrow 0} \Tr\(\gamma\frac{M^2}{D^2 + M^2}\),
\eea
which is the regularization that Callias used. We think that using the more convergent regularization of an exponential function is interesting in itself, as it could possibly extend the Callias index theorem to a wider class of index problems. Therefore we will devote the first part of this Letter on this subject. But let us at once say that our regulator probably has no advantages when attacking these old problems. It does not provide us with a solution for how to count the number of zero modes in a multimonopole configuration with a non-maximally broken gauge group, where the index can not be reliable computed due to a contribution from the continuum portion of the spectrum. What we hope though, is that our regulatization can be useful when attacking our new index problem associated with the moduli space of self-dual strings.

In section \ref{callias} we obtain the index in one and three dimensions. In three dimensions we apply this on the multimonopole moduli space and re-derive the result in \cite{Weinberg}. A recent review article on monopoles and supersymmetry is \cite{Yi}. The one and three-dimensional index problems have also been studied in \cite{Hirayama}. We then indicate how our method manages to reproduce the correct results in any odd dimensions. In section \ref{selfdual} we show how one at least in principle should be able to compute the dimension of the moduli space of $N$ self-dual strings by computing a certain index.

\section{Computing the Callias index in odd-dimensional spaces}\label{callias}
For Dirac operators on open $n-1$-dimensional space where $n-1$ is odd, there is an index theorem by Callias \cite{Callias}. This applies to Dirac equations of the form
\bea
D\psi = 0
\eea
where the Dirac operator $D$ is of the form
\bea
D=\gamma_i iD_i + \gamma_n \phi.
\eea
Here $i=1,...,n-1$ and $\gamma_{\mu}\equiv(\gamma_i,\gamma_n)$ denote the Dirac gamma matrices,
\bea
\{\gamma_{\mu},\gamma_{\nu}\}=2\delta_{\mu\nu}.
\eea
We define the gauge covariant derivative as $iD_{is} = i\partial_{is} + A_{is}$ and all our fields are hermitian. If $n-1$ is odd, the gamma matrices can be represented as
\bea
\gamma_i = \(\begin{array}{cc}
0 & \gamma_i\\
\gamma_i & 0
\end{array}\),\cr
\gamma_n = \(\begin{array}{cc}
0 & i\\
-i & 0
\end{array}\)
\eea
One may use the $n$-dimesional notation $A_{\mu} = (A_i,\phi)$, $D=\gamma_{\mu}iD_{\mu}$, but one must then remember that space is really $n-1$ dimensional.

If $n-1$ is even there is no Weyl representation of the gamma matrices (because of the inclusion of the `gamma-five'), and no index theorem of this form exists. 

We define the `gamma-five' for even $n$ as
\bea
\gamma \equiv -i^{-\frac{n}{2}}\gamma_{1\cdots n}
\eea
which then is hermitian, and we define the projectors
\bea
P_{\pm} = \frac{1}{2}\(1\mp \gamma\).
\eea
In odd dimensions $n-1$, the Dirac operator splits into two Weyl operators
\bea
\D &\equiv& P_+ D P_-\cr
\D^{\dag} &\equiv& P_- D P_+
\eea
Because $P_{\pm}$ and $D$ are all hermitian, it follows that $\D^{\dag}$ is the hermitian conjugate of $\D$. Also, because $D$ is already of an off-block diagonal form, it suffices to include just one of the projectors, so we can just as well write this as
\bea
\D = P_+ D = D P_-\cr
\D^{\dag} = P_- D = D P_+
\eea
The index can now be defined as
\bea
\dim \ker \D - \dim \ker \D^{\dag}
\eea
Since $\ker \D = \ker \(\D^{\dag}\D\)$ and $\ker \D^{\dag} = \ker \(\D\D^{\dag}\)$ we can express this as\footnote{To see this that $\ker \D = \ker \D^{\dag}\D$ we apply the definition of hermitian conjugate with respect to the inner product $(\psi,\chi)=\int dx\psi^{\dag}\chi$ and the property of the norm, to $0=(\psi,\D^{\dag}\D\psi)=(\D\psi,\D\psi)$.}
\bea
\dim \ker \(\D^{\dag}\D\) - \dim \ker \(\D\D^{\dag}\)=\dim \ker \(\gamma D^2\).
\eea
where we have noted that $\gamma = P_- - P_+$.

Callias, Weinberg and others used the regulator 
\bea
I(M^2) = \Tr \(\gamma \frac{M^2}{D^2+M^2}\)\label{Cal}
\eea
to obtain the index as the limit $M^2 \rightarrow 0$. In this Letter we will be slightly more general. We define
\bea
J_i(x,y) \equiv \tr\<x\left|\gamma\gamma_if(D)\right|y\>,
\eea
for any function $f$ (and of course $D$ is not dimensionless, so $D$ has to be accompanied by $M$ in a suitable way). Then we notice that
\bea
W(x,y) &\equiv &\(i\gamma_i \partial_{x^i} + \gamma_{\mu} A_{\mu}(x) + M\)\<x\left|f\(D\)\right|y\>\cr
&=& \<x\left|f\(D\)\right|y\>\(-i\gamma_i\partial_{y^i} + \gamma_{\mu} A_{\mu}(y) + M\)
\eea
where (manifestly)
\bea
W(x,y) = \<x\left|(D+M)f\(D\)\right|y\>.
\eea
From this, we obtain the following identity
\bea
i\(\partial_{x^i} + \partial_{y^i}\)J_i(x,y) &=& 2\tr\<x\left|\gamma Df(D)\right|y\>\cr
&&+\tr\(A_{\mu}(y)-A_{\mu}(x)\)\<x\left|\gamma\gamma_{\mu} f(D)\right|y\>
\eea
In odd dimensions, the second term in the right hand side vanishes as $x$ approaches $y$. This can be seen as being equivalent to the statement that there is no chiral anomaly in odd dimensions (by using point-splitting and inserting a Wilson line). So we get
\bea
i\partial_i J_i(x,x) &=& 2\tr\<x\left|\gamma Df(D)\right|x\>
\eea
If we wish to compute the index as in Eq (\ref{Cal}), then we can take
\bea
f(D) = \frac{1}{D} \frac{M^2}{D^2+M^2}
\eea
(however there is no unique choice of $J_i$). We then get
\bea
J_i(x,y)&=&\tr\<x\left|\gamma\gamma_i\frac{1}{D} \frac{M^2}{D^2+M^2}\right|y\>\cr
&=&\tr\<x\left|\gamma\gamma_i\frac{1}{D} \(-D^2+D^2+M^2\)\frac{1}{D^2+M^2}\right|y\>\cr
&=&-\tr\<x\left|\gamma\gamma_i D \frac{1}{D^2+M^2}\right|y\>.\label{minus}
\eea
provided
\bea
\tr\<x\left|\gamma\gamma_i \frac{1}{D}\right|y\> = 0
\eea
We will see in the next few paragraphs how one can achieve this by using a principal value prescription.

The virtue of expressing Eq (\ref{Cal}) as a total divergence, is that we then can compute the index as a boundary integral over an $(n-2)$-sphere at infinity as
\bea
I(M^2)=\frac{i}{2}\int_{S^{n-2}_{\infty}} d\Omega_{n-2} r^{n-2}\hat{x}_i J_i(x,x).
\eea
where $r$ is the radius of the sphere and $d\Omega_{n-2}$ denotes the volume element of the unit sphere. 

If instead we wish to compute the index as the limit of 
\bea
I(s)=\Tr\(\gamma e^{-sD^2}\).
\eea
as $s\rightarrow \infty$, then we get
\bea
J_i(x,y)&=&\tr\<x\left|\gamma\gamma_i \frac{1}{D} e^{-sD^2}\right|y\>.
\eea
It might seem confusing that we can have a plus sign here, when we have a minus sign in Eq (\ref{minus}). These peculiar signs seem to be correct though. Why we can have opposite signs should be a reflection of the fact that these expressions can not be continuously connected with each other, at least not in any obvious way (like taking $M$ to zero and $s$ to zero. In fact $s$ should be taken to plus infinity as $M$ goes to zero). We will now illustrate how one can use this $J_i$ to compute the index in odd dimensions.

\subsection*{One dimension}
We choose our gamma matrices as
\bea
\gamma_1 = \(\begin{array}{cc}
0 & 1\cr
1 & 0
\end{array}\),
\gamma_2 = \(\begin{array}{cc}
0 & i\cr
-i & 0
\end{array}\)
\eea
and we have
\bea
\gamma = i\gamma_1\gamma_2 = \(\begin{array}{cc}
1 & 0\cr
0 & -1
\end{array}\).
\eea
The Dirac operator reads
\bea
D=i\gamma_1 \partial + \gamma_2 \phi
\eea
We need the square of the Dirac operator,
\bea
D^2 = -\partial^2 + \phi^2 + \gamma \partial \phi.
\eea
We make the choice
\bea
J_1(x,y)=-\tr\<x\left|\gamma\gamma_1 D \frac{1}{D^2 + M^2}\right|y\>
\eea
We assume that $\phi(x)$ converges towards some constant values at $x=-\infty$ and $x=+\infty$. That means that we may ignore $\partial \phi(x)$ for sufficiently large $|x|$, where we then get 
\bea
J_1(x,x) = -\tr\(\gamma\gamma_1\gamma_2\)\int_{-\infty}^{\infty} \frac{dk}{2\pi} \frac{\phi}{k^2 + \phi^2 + M^2} = i\frac{\phi}{\sqrt{\phi^2 + M^2}}
\eea
The index is now given by
\bea
\lim_{M\rightarrow 0} \frac{i}{2}\(J_1(+\infty)-J_1(-\infty)\) = \pm 1
\eea
if $\phi$ flips the sign an odd number of times when going from $-\infty$ to $+\infty$, and $0$ otherwise.

If instead we choose
\bea
J(x,y)&=&\tr\<x\left|\gamma\gamma_1 D\frac{1}{D^2} e^{-sD^2}\right|y\>
\eea
then we get 
\bea
J(x,x) = \tr\(\gamma\gamma_1\gamma_2\)\int \frac{dk}{2\pi} \frac{\phi}{k^2 + \phi^2}e^{-s(k^2 + \phi^2)}
\eea
If we compute the integral over $k$ in the most natural way, then we get a result that vanishes in the limit $s\rightarrow \infty$. Could there be another way of defining this integral, such that we do not get zero as the result? We notice that the integral 
\bea
A(s)\equiv \int dk \frac{e^{-s(k^2+1)}}{k^2 + 1}
\eea
for $s>0$ is convergent only if we integrate $k$ along a line in the complex plane which is such that it asymptotically is such that $-\frac{\pi}{2}<\theta<\frac{\pi}{2}$ where $k=|k|e^{i\theta}$. Integrating along any such line in the complex plane, we get the same value of this integral. If on the other hand we integrate over a line that asymptotically lies outside this cone, then we get a divergent integral for $s>0$. But we get a convergent integral for $s<0$. We then define the value of the integral for $s>0$ as the analytic continuation of the same integral for $s<0$. It remains to compute this convergent integral. Replacing $k$ by $ik$ and $s$ by $-s$, we get the integral
\bea
A(-s)=-i\int_{-\infty}^{\infty} dk \frac{e^{-s(k^2-1)}}{k^2 - 1}
\eea
We can compute its derivative
\bea
A'(-s)=-i\int_{-\infty}^{\infty} dk e^{-s(k^2-1)} = -i\sqrt{\frac{\pi}{s}}e^{s}
\eea
The right-hand side can obviously be analytically continued to $-s$, and that is how we will define $A(s)$ where the integral representation does not converge. We can then integrate up $A'(s)$,
\bea
A(\infty)=A(0)-\int_o^{\infty} ds \sqrt{\frac{\pi}{s}}e^{-s} = A(0) - \sqrt{\pi}\Gamma\(\frac{1}{2}\) = A(0) - \pi
\eea
and we then need to compute 
\bea
A(0) &=& i\int_{-\infty}^{\infty} dk \frac{1}{k^2 - 1}
\eea
We define this as the principal value. This is ad hoc -- we have no argument why one should define it like this. But if we accept this, then we get $A(0)=0$. We conclude that we could just as well define the integral that we had, as
\bea
\lim_{s\rightarrow \infty}\int dk \frac{e^{-s(k^2+1)}}{k^2 + 1} = -\pi.
\eea
But this requires us to perform the integration of $k$ in the cone where it diverges for $s>0$, and then define this integral by analytic continuation. This seem to be rather ad hoc. We have three rather week arguments why one should Wick rotate. First, if we keep $x-y$ as a small number, then we get the factor $e^{ik(x-y)}$ and this can act as a convergence factor only if we Wick rotate. (We illustrate this in the Appendix where we compute the corresponding integral in any complex number of dimensions.) Second, it seems to be the only way that we could produce a non-trivial answer. Third, with this prescription we will manage to reproduce the right answer in any odd number of dimensions, where we can check our result against the safer regularization used by Callias.

If we compute the integral by this prescription, then we get
\bea
J(x,x) = \tr\(\gamma\gamma_1\gamma_2\)\lim_{s\rightarrow \infty} \int \frac{dk}{2\pi} \frac{\phi}{k^2 + \phi^2}e^{-s(k^2+\phi^2)} = i \frac{\phi}{\sqrt{\phi^2}}
\eea
and we see that we indeed get the right answer.

\subsection*{Three dimensions and magnetic monopoles}
The physics problem that we will consider in three dimensions, is to compute number of zero modes of the Bogomolnyi equation 
\bea
F_{ij}=\epsilon_{ijk}D_k\phi
\eea
We choose the convention that our fields are hermitian. It is convenient to group the fields into `gauge potential'
\bea
A_{\mu}=(A_i,\phi)
\eea
We define $D_{\mu}=(D_i,\phi)$ such that $iD_{\mu}=i\partial_{\mu}+A_{\mu}$ and we let $G_{\mu\nu}=i[D_{\mu},D_{\nu}]$ be the associated `field strength'. Then the Bogomolnyi equation reads
\bea
G_{\mu\nu} =\frac{1}{2}\epsilon_{\mu\nu\rho\sigma}G_{\rho\sigma}.
\eea
Linearizing this, we get
\bea
D_{\mu}\delta A_{\nu} = \frac{1}{2}\epsilon_{\mu\nu\rho\sigma}D_{\rho}\delta A_{\sigma}
\eea
Contracting with $\gamma_{\mu\nu}$, we get
\bea
(1+\gamma)\gamma_{\mu\nu}D_{\mu}\delta A_{\nu} = 0
\eea
and if we impose the background gauge condition
\bea
D_{\mu}\delta A_{\mu} = 0
\eea
which is to say that zero modes are orthogonal to gauge variations with respect to the moduli space metric, then we can write this linearized equation as a Dirac equation
\bea
D\psi\equiv \gamma_{\mu}D_{\mu}\psi = 0
\eea
where
\bea
\psi := (1+\gamma)\gamma_{\mu}\delta A_{\mu}.
\eea
We compute
\bea
D^2 = -D_i^2 + \phi^2 + \frac{1}{2}i\gamma_{\mu\nu}G_{\mu\nu}
\eea
Inserting the Bogomolnyi configuration we can write this, thus using the fact that $G_{\mu\nu}$ is selfdual, 
\bea
D^2 = -D_i^2 + \phi^2 + \frac{1}{4}(1+\gamma)i\gamma_{\mu\nu}G_{\mu\nu}.
\eea
and get a vanishing theorem. Namely, $\dim \ker \D\D^{\dag} = 0$ as $\D\D^{\dag}>0$ is strictly postive. Hence we can compute the dimension of the moduli space $\dim \ker \D \equiv \dim \ker \D^{\dag}\D$ just by computing the index of $\D$. To compute the index, we now wish to compute
\bea
J_i(x,x) = \tr\<x\left|\gamma\gamma_i\gamma_k D_k \frac{1}{D^2} e^{-s D^2}\right|x\>
\eea
We assume that asymptotically $\phi$ approaches a constant value at infinity. This corresponds to a gauge choice where we have a Dirac string singularity. Some further examination reveals that we get a non-negligible contribution to $J_i$, for a sufficiently large two-sphere, only from the term
\bea
J_i(x,x) = \tr\(\gamma\gamma_i\gamma_4 \phi\int \frac{d^3 k}{(2\pi)^3} \frac{1}{k^2 + \phi^2 + \frac{1}{2}i\gamma_{\mu\nu}G_{\mu\nu}} e^{-s\(k^2+ \phi^2 + \frac{1}{2}i\gamma_{\mu\nu}G_{\mu\nu}\)}\)
\eea
We thus need to perform an integral of the form 
\bea
A(s)=\int dk \frac{k^2}{k^2+1} e^{-s(k^2+1)}
\eea
If we choose the same prescription as we did in one dimension, then we get the result
\bea
A(+\infty)=\pi.
\eea
For details of such a computation we refer to appendix $A$.

If we apply this result to the integral that we had, we get
\bea
J_i(x,x) = \frac{1}{2\pi}\tr\(\gamma\gamma_i\gamma_4 \phi\sqrt{\phi^2 + \frac{1}{2}i\gamma_{\mu\nu}G_{\mu\nu}}\)
\eea
We expand the square root,
\bea
\sqrt{\phi^2 + \frac{1}{2}i\gamma_{\mu\nu}G_{\mu\nu}} = \phi + \frac{1}{4\sqrt{\phi^2}}i\gamma_{\mu\nu}G_{\mu\nu} + ...
\eea
In the far distance, in a charge $Q$ monopole configuration, we find that
\bea
\gamma_{\mu\nu}G_{\mu\nu} = 2\gamma_k\gamma_4(1-\gamma)\frac{\hat{x}_k}{r^2}Q
\eea
and so when we trace over the gamma matrices, we get
\bea
J_i(x,x)=\frac{i\hat{x}^i}{2\pi r^2}\tr\(\frac{\phi Q}{\sqrt{\phi^2}}\).
\eea
If we now for instance assume $SU(2)$ gauge group, broken to $U(1)$, then if we integrate $\frac{i}{2}J_i$ over $S^2$, we get the index $2Q$. The number of bosonic zero modes is twice the index, i.e. $-4Q$ in our conventions \cite{Weinberg,Yi}.

\subsection*{$(2m+1)$ dimensions}
In $2m+1$ dimensions we get the integral 
\bea
A(\mu)\equiv \lim_{s\rightarrow \infty}\int dk \frac{k^{2m}}{k^2+\mu^2} e^{-s(k^2+\mu^2)}
\eea
if we use our regulator. Here
\bea
\mu^2 \equiv v^2 + G
\eea
(and $G$ is an abbreviation for $\frac{1}{2}i\gamma_{\mu\nu}G_{\mu\nu}$.) This should be compared to the integral
\bea
B(\mu)\equiv -\lim_{M\rightarrow 0}(-1)^m\int dk \frac{k^{2m}}{\(k^2 + v^2 + M^2\)^{m+1}}G^m
\eea
that we get using the Callias regulator. \footnote{This integral comes from expanding 
\bea
\frac{1}{k^2 + v^2 + G + M^2} = \frac{1}{k^2+v^2+M^2} + ...
\eea
in powers of $G$ as a geometric series \cite{Weinberg}.} In order to compare these integrals, we rewrite them as
\bea
A(\mu) &=& \mu^{2m-1}a\cr
B(\mu) &=& v^{-1}b G^m
\eea
where
\bea
a &=& \lim_{\t s\rightarrow \infty}\int d\xi \frac{\xi^{2m}}{\xi^2 + 1} e^{-\t s\(\xi^2+1\)}\cr
b &=& -\lim_{\t M\rightarrow 0}(-1)^m\int d\xi \frac{\xi^{2m}}{\(\xi^2+1+\t M^2\)^{m+1}}
\eea
We compute $a$ according the prescription introduced above in one and three dimensions, that is by Wick rotating $\xi$ and continue analytically in $s$. (Details are in appendix $A$.) We can compute $b$ using residue calculus (introducing a regulator so that we can close the contour on a semi-circle at infinity). The result is
\bea
a &=& -(-1)^m\pi\cr
b &=& (-1)^m\frac{1}{2}\pi\frac{\Gamma\(m-\frac{1}{2}\)}{\Gamma\(\frac{1}{2}\)}
\eea
We next expand 
\bea
v A(\mu) &=& v\(v^2 + G\)^{m-\frac{1}{2}}a\cr
&=& v^{2m}a + ...+\frac{\Gamma\(m-\frac{1}{2}\)}{\Gamma\(-\frac{1}{2}\)}a G^m + ...\cr
v B(\mu) &=& b G^m
\eea
and we find that the coefficient of $G^m$ becomes equal to
\bea
-(-1)^m\frac{\Gamma\(m-\frac{1}{2}\)}{\Gamma\(-\frac{1}{2}\)}\pi
\eea 
if one uses our regularization, and equal to
\bea
(-1)^m\frac{1}{2}\frac{\Gamma\(m-\frac{1}{2}\)}{\Gamma\(\frac{1}{2}\)}\pi
\eea
if one uses the Callias regularization. We see that the two expressions coincide for all $m$.

We have now showed that if we use our prescription of Wick rotating $k$ to compute the integrals over the exponential, then we get the right answer for all cases that can be safely computed using a regulator that is less convergent. We are inclined to think that our prescription for how to compute the integral, will also work for index problems where the Callias regulator diverges. But we have no proof. It is perhaps not so obvious that more general index problems can be formulated. In the next section we will give one example of a more general type of index problem.

\section{Four dimensions and self-dual strings}\label{selfdual}
To introduce the notation, we first consider the free Abelian tensor multiplet theory in $1+5$ dimensions. The on-shell field content is a two-form gauge potential $B_{\mu\nu}$, five scalar fields $\phi^A$ and corresponding Weyl fermions $\psi$. The field strength $H_{\mu\nu\rho} = \partial_{\mu}B_{\nu\rho}+\partial_{\rho}B_{\mu\nu}+\partial_{\nu}B_{\rho\mu}$ is selfdual. The supersymmetry variation of the Weyl fermions is
\bea
\delta \psi = \(\frac{1}{12}\Gamma^{\mu\nu\rho}H_{\mu\nu\rho} + \Gamma^{\mu}\Gamma_A \partial_{\mu} \phi^A\)\epsilon
\eea
where we use eleven-dimensional gamma matrices splitted into $SO(1,5)\times SO(5)$, so that in particular
\bea
\{\Gamma^{\mu},\Gamma_A\} &=& 0.
\eea
In a static and $x^5$ independent field configuration, in which only $\phi^5 =: \phi$ is non-zero, we find the SUSY variation
\bea
\delta \psi = \(\Gamma^{0i5}H_{0i5} + \Gamma^{i}\Gamma_{A=5} \partial_i \phi\)\epsilon
\eea
If we assume that the classical bosonic field configuration is such that
\bea
\partial_i\phi = H_{0i5} 
\eea
then the SUSY variation reduces to
\bea
\delta \psi = \partial_i \phi \Gamma^i\(\Gamma^{05} + \Gamma_{A=5}\) \epsilon
\eea
and we find the condition for unbroken SUSY as
\bea
\(1 + \Gamma^{05}\Gamma_{A=5}\) \epsilon = 0
\eea
If we use the Weyl condition
\bea
\Gamma \epsilon = -\epsilon
\eea
of the $(2,0)$ supersymmetry parameter $\epsilon$, then we can also write this as
\bea
\(1 + \Gamma^{1234}\Gamma_{A=5}\) \epsilon = 0.
\eea
We may represent the gamma matrices as
\bea
\Gamma_{\mu} &=& \(\Gamma_0,\Gamma_i,\Gamma_5\) = \(1\otimes i\sigma^2\otimes 1,\gamma_i\otimes \sigma^1\otimes 1,\gamma\otimes \sigma^1\otimes 1\)\cr
\Gamma_A &=& 1\otimes i\sigma^2 \otimes \sigma_A
\eea
where $\sigma^{1,2,3}$ are the Pauli sigma matrices, $\gamma=\gamma_{1234}$. Then the condition for unbroken SUSY is
\bea
\(1+\gamma \otimes \sigma\) \epsilon = 0
\eea
where $\sigma = \sigma_{1234} = \sigma_{A=5}$.

We have found that if
\bea
H_{ijk} = \epsilon_{ijkl}\partial_l \phi
\eea
then half SUSY is unbroken. This equation is the Bogomolnyi equation for self-dual strings \cite{Lambert}. We are interested in finding the number of parameters needed to describe solutions of this equation. We can linearize it and get the equation
\bea
\gamma_i \partial_i \chi = 0\label{line}
\eea
for the bosonic zero modes, that we have gathered into a matrix
\bea
\chi \equiv \gamma_{ij}\delta B_{ij} + \gamma \delta \phi.
\eea
For this to work we must also assume the background gauge condition 
\bea
\partial_i B_{ij} = 0.
\eea
Now this linearized equation Eq (\ref{line}) does not make any reference to the gauge field. So there is no way that we could count the number of parameters of a multi-string configuration just using this equation. This should of course not be a surprise. The strings that we have in the Abelian theory are not solutions of the field equations. They have to be inserted by hand, that is we need to insert delta function sources by hand, in the same spirit as for Dirac monopoles. 

To be able to count the number of zero modes, we must consider some interacting theory which (at the classical level) has solitonic string solutions.

To pass to non-Abelian theory we begin by rewriting the Abelian theory in loop space. Loop space consists of parametrized loops $C$: $s\mapsto C^{\mu}(s)$. We introduce the Abelian `loop fields' \cite{Gust}
\bea
A_{\mu s} &=& B_{\mu\nu}(C(s))\dot{C}^{\nu}(s)\cr
\phi^{\mu s}&=& \phi(C(s))\dot{C}^{\mu}(s)\cr
\psi^{\mu s}&=& \psi(C(s))\dot{C}^{\mu}(s)
\eea
With these definitions, a short computation reveals that $A_{\mu s}$ transforms as a vector and $\phi^{\mu s}$ a contra-variant vector under diffeomorphisms in loop space induced by diffeomorphisms in space-time. One may then extend these transformation properties to any diffeomorphism in loop space. Space-time diffeomorphism and reparametrizations of the loops then get unified and are both diffemorphisms in loop space. The only thing to remember is what is kept fixed under the variation. If it is the parameter of the loop, or the loop itself. 

The field strength becomes
\bea
F_{\mu s,\nu t} = H_{\mu\nu\rho}(C(s))\dot{C}^{\rho}(s)\delta(s-t)
\eea
In terms of these fields, the Bogomolnyi equation will read\footnote{We denote by $\partial_{is}$ the usual functional derivative with respect to $C^{\mu}(s)$.}
\bea
F_{is,jt} = \epsilon_{ijkl}\partial_{k(s}\phi_{lt)}.
\eea

We pass to the non-Abelian theory by letting these loop fields become non-Abelian, in the sense that $A_{\mu s} = A^a_{\mu s}\lambda^a(s)$ where $\lambda^a(s)$ are generators of a loop algebra associated to the gauge group \cite{Gust}. We introduce a covariant derivative
\bea
D_{\mu s} &=& \partial_{\mu s} + A_{\mu s}.
\eea
Local gauge transformations act as
\bea
\delta_{\Lambda} A_{\mu s} &=& D_{\mu s}\Lambda\cr
\delta_{\Lambda} \phi^{\mu s} &=& [\phi^{\mu s},\Lambda].
\eea

Given a loop $C$, we automatically get a tangent vector $\dot{C}^{\mu}(s)$ that makes no reference to space-time. We can therefore impose the loop space constraints
\bea
\dot{C}^{\mu}(s) A_{\mu s} = 0
\eea
for each $s$, and also
\bea
\phi^{\mu s} = \dot{C}^{\mu}(s) \phi(s;C)
\eea
for some subtle field $\phi(s;C)$ on loop space. As a consequence, we find that
\bea
A_{\mu s}\phi^{\mu s} = 0.
\eea
These constraints are covariant under diffeomorphisms of space-time and reparametrizations of loops. They are invariant also under local gauge transformations, provided that the gauge parameter is subject to the condition
\bea
\dot{C}^{\mu}(s)\partial_{\mu s} \Lambda = 0
\eea
which is the condition of reparametrization invariance. With the assumption made that $\lambda^a(s)$ are generators of a loop algebra, we find that the constraint can also be written as 
\bea
[A_{\mu s},\phi^{\mu t}] = 0\label{fr}
\eea
A local gauge variation of this constraint is
\bea
[D_{\mu s}\Lambda, \phi^{\mu t}] + [A_{\mu s},[\phi^{\mu t},\Lambda]] 
&=& [\partial_{\mu s}\Lambda,\phi^{\mu t}] + [[A_{\mu s},\Lambda],\phi^{\mu t}] + [A_{\mu s},[\phi^{\mu t},\Lambda]]\cr
&=& [\partial_{\mu s}\Lambda,\phi^{\mu t}] + [\Lambda,[\phi^{\mu t},A_{\mu s}]]
\eea
The last term vanishes by the constraint. The first term gives us the constraint Eq (\ref{fr}) that we must impose on the gauge parameter
\bea
\Lambda = \int ds \Lambda^a(s,C)\lambda^a(s).
\eea

We have now introduced non-local non-Abelian fields with infinitely many components. It is also likely that consisteny of the theory requires an infinite set of constraints on these fields. Maybe then, it could be that we may in the end descend to a finite degrees of freedom. But this is just a speculation. The problem appears to be difficult and ill-defined -- How should one define a degree of freedom in a strongly coupled non-local theory?

The non-Abelian generalization of the Bogomolnyi equation should be given by \cite{Gust}
\bea
F_{is,jt} = \pm \epsilon_{ijkl}D_{k(s}\phi_{lt)}.
\eea
This equation is gauge invariant and invariant under the residual $SO(4)$ Lorentz group that is preserved by the strings. We can not think of any reasonable modification of this equation that would preserve these symmetries, so on this grounds alone one could suspect  this equation to be correct. Of course this is not the only requirement that the BPS condition imposes. We also get conditions on the $0s$ and the $5s$ components. But these BPS equations will be of no interest to us right now.

We will show below that the linearized Bogomolnyi equation can be written as
\bea
\gamma_i \(D_{i(s} + \sigma \phi_{i(s} \) \chi_{t)} = 0\label{linn}
\eea
We will also see below that we (presumably) can actually drop the symmetrization in $s$ and $t$ in this equation. The fields transform in the adjoint representation of the loop algebra, by which we mean that $\phi_{is}\chi_t = [\phi_{is},\chi_t]$. We define the Dirac operator 
\bea
D_s = \gamma_i \(D_{is} + \sigma \phi_{is} \)
\eea
and the projectors
\bea
P_{\pm} \equiv \frac{1}{2}\(1\mp \gamma\sigma\),
\eea
We can now formulate an index problem, in an even-dimensional (loop-)space. The even-dimensional space in this case is given by the $4$-dimensional transverse space to the strings, and the index is given by
\bea
\dim \ker \D_s - \dim \ker \D^{\dag}_s
\eea
where
\bea
\D_s = P_+ D_s = D_s P_-\cr
\D^{\dag}_s = P_- D_s = D_s P_+.
\eea
Since $D_s$ and $P_{\pm}$ are hermitian, it is manifest that $\D^{\dag}_s$ defined this way will be the hermitian conjugate of $\D_s$, thus justifying the notation.

Computing the index alone is not sufficient in order to obtain the dimension of the moduli space of self-dual strings. We also need a vanishing theorem that says that $\dim \ker \D^{\dag}_s=0$.

Linearizing the Bogomolnyi equation, we get
\bea
2D_{[is}\delta A_{jt]} = \pm\epsilon_{ijkl}\(D_{ks}\delta \phi_{lt} + \phi_{ks}\delta A_{lt}\)
\eea
Contracting by $\gamma^{ij}$, we get
\bea
\gamma^{ij}\t D_{is}\chi_{jt}=0
\eea
where we have defined
\bea
\t D_{is} &\equiv & D_{is} \mp \gamma \phi_{is}\cr
\chi_{is} &\equiv & \delta A_{is} \mp \gamma \delta \phi_{is}
\eea
To see that the linearized BPS equation can be written like this, one must use the constraint
\bea
\gamma^{ij}\phi_{is}\delta\phi_{jt}=0.
\eea
We can avoid having explicit $\pm$ signs by introducing the other chiraly matrix at our disposal, namely $\sigma$ that lives in a different vector space than $\gamma$. We can then hide the $\pm$ signs in the tensor product  
\bea
\gamma \otimes\sigma = \pm 1 
\eea
which amounts to 
\bea
\t D_{is} &\equiv & D_{is} + \sigma \phi_{is}\cr
\chi_{is} &\equiv & \delta A_{is} + \sigma \delta \phi_{is}
\eea
without any $\pm$.\footnote{To really understand what is going on, one should apply $\(1\pm \gamma\sigma\)$ on everything, on $\psi_s$ and on $D_s$. Then one notices that
\bea
\mp \gamma\(1\mp \gamma\sigma\) = \sigma\(1 \mp \gamma\sigma\).
\eea
That is, we can trade $\mp \gamma$ for $\sigma$, once we apply $\(1\pm \gamma\sigma\)$ on everything. This is what we really should do, but to keep the notation simple, we do not spell this out.}
If we define
\bea
\chi_s \equiv \gamma^i \chi_{is}
\eea
then we can write the zero mode equation as
\bea
\gamma^{i}\t D_{is}\chi_t + \t D^{i}_s \chi_{it} = 0.
\eea
Let us analyze the second term in this equation. It is given by
\bea
&&D^i_s\delta A_{it} + \phi^i_s \delta\phi_{it}\cr
&-&\gamma\(\phi^i_s\delta A_{it} + D^{i}_s \delta \phi_{it}\)
\eea
We should not count variations that are gauge variations as bosonic zero modes. We can insure this by demanding the zero modes to be orthogonal to gauge variations, with respect to the metric on the moduli space,
\bea
\(\delta_{\Lambda}A_{is},\delta A_{it}\)+\(\delta_{\Lambda}\phi_{is},\delta \phi_{jt}\) = 0
\eea
This leads to the background gauge condition 
\bea
D^i_s\delta A_{it} + \phi^i_s \delta\phi_{it} = 0.
\eea
This condition implies that the gauge variation of the zero modes vanishes,
\bea
\delta_{\Lambda}\delta A_{is} = 0 = \delta_{\Lambda}\delta\phi_{is}
\eea
To see this, we make a gauge variation $\delta_{\Lambda} \delta A_{is} = D_{is}\Lambda$, $\delta_{\Lambda} \phi_{is} = \phi_{is}\Lambda$, and ask which gauge parameters $\Lambda$ will respect the background gauge condition. Inserting this gauge variation into the background gauge condition, we get
\bea
\(D^i_s D_{it} + \phi^i_s \phi_{it}\) \Lambda = 0.
\eea
For this to work nicely, it seems that we must constrain the non-locality of our loop field such that $\partial^i_{(s}\partial_{it)}<0$. Then the only solution to this equation is $\Lambda = 0$. In other words all gauge variations of the zero modes have to vanish.
 
Furthermore we want the variation to preserve the orthogonality between $A_{is}$ and $\phi_{is}$,
\bea
\(A_{is},\delta \phi_{it}\) + \(\delta A_{is},\phi_{it}\) = 0
\eea
If we make a gauge variation of this, then we get the condition 
\bea
\(\delta_{\Lambda}A_{is},\delta \phi_{it}\) + \(\delta A_{is},\delta_{\Lambda}\phi_{it}\) = 0
\eea
which amounts to
\bea
\phi^i_s\delta A_{it} + D^{i}_s \delta \phi_{it} = 0.
\eea

We conclude that the zero mode equation can be written as
\bea
D_{s}\chi_t = 0
\eea
where 
\bea
D_s = \gamma_i \(D_{is} + \sigma \phi_{is}\)
\eea
We are interested in counting the number of such modes in a background of $k$ BPS strings. We compute
\bea
D^2 &=& (D_{is})^2 + (\phi_{is})^2 + \frac{1}{2}\gamma^{ij} \(F_{is,js} + \gamma\sigma \epsilon_{ijkl} D_{ks}\phi_{ls}\)
\eea
(Here $D^2\equiv D_sD_s\equiv \int \frac{ds}{2\pi} D_{s}D_{s}$, and analogously for the other fields or operators.) In a BPS configuration, we get is
\bea
D^2 &=& (D_{is})^2 + (\phi_{is})^2 + \frac{1}{2}\gamma^{ij}\(1+\gamma\sigma\) F_{is,js}
\eea
Furthermore, in the subspace where $1+\gamma\sigma=0$, we find that
\bea
D^2 &=& (D_{is})^2 + (\phi_{is})^2
\eea
is a strictly negative operator, hence has no zero modes. This means that we have a vanishing theorem, $\dim \ker \D^{\dag} = 0$.

\subsection*{A small comment}
The zero mode equation was really 
\bea
D_{(s}\chi_{t)} = 0\label{qq}
\eea 
where we should symmetrize in $s$ and $t$. That means that we should rather consider
\bea
D_s D_{(s}\chi_{t)} &=& \frac{1}{2} \(D_s D_s\chi_t + D_s D_t \chi_s\)\cr
&=& \frac{1}{2} \(D_s D_s\chi_t +  D_t D_s \chi_s + [D_s,D_t]\chi_s\).
\eea
If now $D_{[s}D_{t]}=0$ and $D_s\chi_s=0$, then we get 
\bea
D_s D_s \chi_t=0
\eea
The latter condition, $D_s\chi_s=0$ is of course a consequence of $D_{(s}\chi_{t)} = 0$ with $s=t$. The former condition reads
\bea
0 &=& D_{[s}D_{t]}\cr
&=& D_{i[s}D_{it]} + \phi_{i[s}\phi_{it]} + \sigma D_{i[s}\phi_{it]}
\eea
which we would like to impose as a constraint. Restricting to the abelian case this is condition is of course true as $0\equiv \partial_{i[s}\partial_{|i|t]}$. If we can impose this as a constraint on the non-abelian fields, then we have now seen that the zero mode equation Eq (\ref{qq}) implies that
\bea
\int ds D_s^{\dag}D_s \chi_t=0\label{ww}
\eea
because $D_s$ is anti-self-adjoint with respect to the inner product 
\bea
\(\psi_s,\chi_t\)=\int \D C \tr\(\psi^{\dag}_s(C)\chi_t(C)\)
\eea
on loop space. We can also go in the opposite direction. Assuming that Eq (\ref{ww}) holds, we get
\bea
0 = \(\chi_t,D_s^{\dag} D_s \chi_t\) = \(D_s\chi_t,D_s \chi_t\)
\eea
and we conclude that (\ref{qq}) implies 
\bea
D_s\chi_t = 0
\eea
with no symmetrization in $s,t$.

\subsection*{How to compute the index}
We should now be able to compute an index associated to self-dual strings, as the limit
\bea
I(s)=\Tr \(\gamma\sigma e^{sD^2}\)
\eea
when $s\rightarrow \infty$. We define the quantity
\bea
J_{is}(C,C') = \tr\<C\left|\gamma\sigma \gamma_i \gamma_k \(D_{ks} + \sigma \phi_{ks}\) \frac{1}{D^2} e^{sD^2}\right|C'\>
\eea
(it should be clear that the two $s$'s involved in this formula are totally unrelated) and find that
\bea
I(s) = \int \frac{ds}{2\pi}\int \D C \partial_{is}J^{is}(C,C)
\eea
We can separate the functional integral over parametrized loops $C$ into several pieces. We can keep a point on the loops $C(s)=x$ fixed, and separate it as
\bea
\int \D C = \int d^4 x \int \D_x C
\eea
Then we can write $I(s)$ as an integral over a large three-sphere at spatial infinity,
\bea
\int \frac{ds}{2\pi} \int d^4 x \int \D_x C \frac{\partial J_{is}(C)}{\partial C^i(s)}
= \int \frac{ds}{2\pi} \int_{S^3} d\Omega_3 \hat{x}^i \int \D_x C J_{is}(C,C)
\eea
where thus $x=C(s)$.

If we assume that the gauge group is maximally broken to a product of $U(1)$'s by the Higgs vacuum expectation values, then we should have $U(1)$ loop fields at spatial infinity.

If we assume that the gauge group is $SU(2)$ and that it is broken to $U(1)$, then we need only the asymptotic form of the $U(1)$ fields at spatial infinity,\bea
F_{is,jt} &=& H_{ijl}(x)\dot{C}^l(s)\delta(s-t)\cr
\phi_{ks} &=& v \dot{C}_k(s)
\eea

Without doing any computations, we can guess what the outcome of the index calculation should be. A term like
\bea
\epsilon_{ijkl}\int \D_x C  \tr\(F_{is,jt}(C)F_{ks,jt}(C)\)
\eea
could certainly arise somewhere (in odd dimensions a corresponding term vanished since there is no chiral anomaly in odd dimensions). In our case this term vanishes identically by the Bogomolnyi equation and the constraint\footnote{For $U(1)$ fields this would read
\bea
F_{is,jt}\partial_{is}\phi_{jt} \sim H_{ijk}(C(s))\partial_i\phi(C(s))\dot{C}^{k}(s)\dot{C}^j(s)\delta(s-t)^2 \equiv 0.
\eea}
\bea
F_{is,jt}D_{is}\phi_{jt} = 0.
\eea
Then there can be a term
\bea
\epsilon_{ijkl}\int \D_x C  \tr\(\frac{F_{is,jt}\phi_{ks}}{v}\)
\eea
that should arise in a very similar way as the corresponding term arose for monopoles. If we insert the asymptotic $U(1)$ fields, this term becomes proportional to
\bea
\epsilon_{ijkl}H_{ijk}(x)
\eea
That means that the index should be given by some numerical constant, times the magnetic charge
\bea
\int_{S^3_{\infty}} H.
\eea

\newpage
\appendix
\section{Integrals over the exponential}
The integral we will analyze here is
\bea
a(s) = \int_{-\infty}^{+\infty} dk \frac{k^{2\zeta}}{k^2+1}e^{-s\(k^2+1\)}e^{i\epsilon k}
\eea
for any complex number $\zeta$. (The $\epsilon>0$, say, will be taken towards zero. It arose from $\epsilon=x-y$ and we keep it here just as a convergence factor.) We first compute
\bea
a(0) = \int_{-\infty}^{+\infty} dk \frac{k^{2\zeta}}{k^2+1}e^{i\epsilon k}
\eea
In order to make this integral converge for any $\zeta$, we should Wick rotate $k$ to $ik$, and henceforth we will always mean by $i$ the branch $e^{i\pi/2}$, and by $-1$ we mean $e^{i\pi}$. Then we get
\bea
a(0) = -i^{2\zeta+1}\int_{-\infty}^{+\infty} dk \frac{k^{2\zeta}}{k^2-1}e^{-\epsilon k}
\eea
and this integral we evaluate as a principal value. That means to evaluate the residues along the real axis and multiply them not by $2\pi i$, but by half of it, that is, by $\pi i$. We get
\bea
a(0) = (-1)^{\zeta}\pi \frac{1-(-1)^{2\zeta}}{2}.\label{er}
\eea

Next we turn to our integral $a(s)$. It is easier to first compute the derivative. We should still work with the Wick rotated integral. Making the substitution $\xi=k^2$ we can put it on the form of two gamma functions. The result is that 
\bea
a'(-s) &=& -e^{i\pi\(\zeta+\frac{1}{2}\)}\frac{1}{2}\(1+(-1)^{2\zeta}\) \Gamma\(\zeta+\frac{1}{2}\) s^{-\zeta-\frac{1}{2}}e^{s}
\eea
which we can trivially continue analytically to $+s$, and then integrate up. The result is
\bea
a(+\infty) = -\pi\frac{1+(-1)^{2\zeta}}{2}\frac{1}{\cos(\pi \zeta)} + \pi(-1)^{\zeta}\frac{1-(-1)^{2\zeta}}{2}.
\eea


\vskip 0.5truecm
\newpage
\begin{thebibliography}{999}
\bibitem{Lambert}
  P.~S.~Howe, N.~D.~Lambert and P.~C.~West,
  ``The self-dual string soliton,''
  Nucl.\ Phys.\  B {\bf 515}, 203 (1998)
  [arXiv:hep-th/9709014].
\bibitem{Harvey}
  D.~S.~Berman and J.~A.~Harvey,
  ``The self-dual string and anomalies in the M5-brane,''
  JHEP {\bf 0411}, 015 (2004)
  [arXiv:hep-th/0408198].
\bibitem{Callias}
  C.~Callias,
  ``Index Theorems On Open Spaces,''
  Commun.\ Math.\ Phys.\  {\bf 62}, 213 (1978).
\bibitem{Weinberg}
  E.~J.~Weinberg,
  ``Parameter Counting For Multi - Monopole Solutions,''
  Phys.\ Rev.\  D {\bf 20}, 936 (1979).
\bibitem{Yi}
  E.~J.~Weinberg and P.~Yi,
  ``Magnetic monopole dynamics, supersymmetry, and duality,''
  Phys.\ Rept.\  {\bf 43}, 65 (2007)
  [arXiv:hep-th/0609055].
\bibitem{Hirayama}
  M.~Hirayama,
  ``Supersymmetric Quantum Mechanics And Index Theorem,''
  Prog.\ Theor.\ Phys.\  {\bf 70}, 1444 (1983).
\bibitem{Gust}
  A.~Gustavsson,
  ``A reparametrization invariant surface ordering,''
  JHEP {\bf 0511}, 035 (2005)
  [arXiv:hep-th/0508243].\\
  A.~Gustavsson,
  ``The non-Abelian tensor multiplet in loop space,''
  JHEP {\bf 0601}, 165 (2006)
  [arXiv:hep-th/0512341].
\end{thebibliography}
\end{document}